\documentclass[reprint,prb,superscriptaddress]{revtex4-1}

\usepackage{upgreek}
\usepackage[latin1]{inputenc}
\usepackage{subsupscripts}
\usepackage{subscript}
\usepackage{epspdfconversion}

\usepackage{graphicx}
\usepackage{dcolumn}
\usepackage{bm}
\usepackage{color}

\newcommand{\mrm}{\mathrm}

\usepackage{booktabs}
\usepackage{amssymb}

\begin{document}

\preprint{}

\title{Experimental realization of a polariton beam amplifier}

\author{Dominik Niemietz}
\affiliation{Experimentelle Physik 2, Technische Universit\"at Dortmund, \mbox{D-44221 Dortmund, Germany}}
\author{Johannes Schmutzler}
\affiliation{Experimentelle Physik 2, Technische Universit\"at Dortmund, \mbox{D-44221 Dortmund, Germany}}
\author{Przemyslaw Lewandowski}
\affiliation{Department of Physics and CeOPP, University of Paderborn, Warburger Str. 100, D-33098 Paderborn, Germany}
\author{Karol Winkler}
\affiliation{Technische Physik, Physikalisches Institut, Wilhelm Conrad R\"ontgen Research Center for Complex Material Systems,
Universit\"at W\"urzburg, D-97074 W\"urzburg, Germany}
\author{Marc A{\ss}mann}
\affiliation{Experimentelle Physik 2, Technische Universit\"at Dortmund, \mbox{D-44221 Dortmund, Germany}}
\author{Stefan Schumacher}
\affiliation{Department of Physics and CeOPP, University of Paderborn, Warburger Str. 100, D-33098 Paderborn, Germany}
\author{Sebastian Brodbeck}
\affiliation{Technische Physik, Physikalisches Institut, Wilhelm Conrad R\"ontgen Research Center for Complex Material Systems,
Universit\"at W\"urzburg, D-97074 W\"urzburg, Germany}
\author{Martin Kamp}
\affiliation{Technische Physik, Physikalisches Institut, Wilhelm Conrad R\"ontgen Research Center for Complex Material Systems,
Universit\"at W\"urzburg, D-97074 W\"urzburg, Germany}
\author{Christian Schneider}
\affiliation{Technische Physik, Physikalisches Institut, Wilhelm Conrad R\"ontgen Research Center for Complex Material Systems,
Universit\"at W\"urzburg, D-97074 W\"urzburg, Germany}
\author{Sven H\"ofling}
\affiliation{Technische Physik, Physikalisches Institut, Wilhelm Conrad R\"ontgen Research Center for Complex Material Systems,
Universit\"at W\"urzburg, D-97074 W\"urzburg, Germany}
\affiliation{SUPA, School of Physics and Astronomy, University of St Andrews, St Andrews, KY16 9SS, United Kingdom}

\author{Manfred Bayer}
\affiliation{Experimentelle Physik 2, Technische Universit\"at Dortmund, \mbox{D-44221 Dortmund, Germany}}
\affiliation{A. F. Ioffe Physical-Technical Institute, Russian Academy of Sciences, St Petersburg 194021, Russia} 

\date{\today}

\begin{abstract} 
In this report we demonstrate a novel concept for a planar cavity polariton beam amplifier using non-resonant excitation. In contrast to resonant excitation schemes, background carriers are injected which form excitons, providing both gain and a repulsive potential for a polariton condensate. Using an attractive potential environment induced by a locally elongated cavity layer, the repulsive potential of the injected background carriers is compensated and a significant amplification of polariton beams is achieved without beam distortion.    
\end{abstract}



\maketitle
\section{Introduction}
Since the first unambiguous demonstration of polariton condensates in semiconductors \cite{Kasprzak2006}, microcavity polaritonics has gained much attention as this field of research is an appealing platform for the realization of artificial functional potentials and band structures \cite{Tosi2012,Assmann2012,Kim2013,Jacqmin2014,Schneider2015}, novel energy efficient light sources \cite{Tsintzos2008,Schneider2013} and logic circuits \cite{Ortega2013,Sturm2014,Gao2012}.
For the latter, significant progress was achieved recently as all-optical, cascadable transistor operation of resonantly excited polariton fluids \cite{Ballarini2013} as well as the realization of waveguides by background carrier injection using non-resonant optical excitation \cite{Schmutzler2015} were demonstrated. However, a further requirement is the possibility to amplify polaritons to allow for signal transmission over macroscopic distances as the invention of erbium-doped fiber amplifiers has demonstrated long range communication using optical fibers \cite{Desurvire1987}.

A convenient way to realize amplification of propagating polaritons lies in blue-detuned pumping of large sample areas with excitation power below the bistability threshold in combination with a local injection of a polariton fluid on the upper branch of the bistability curve \cite{Amo2010a,Adrados2011,Sich2012}. However, this approach requires rather large laser power, a careful selection of excitation angle and energy and is not feasible for electrical pumping. Another possibility for amplification is the injection of background carriers which act as gain medium for a polariton condensate \cite{Wertz2012}. Moreover, background carriers also provide a repulsive potential mediated by Coulomb interaction causing the deflection of a polariton condensate. On the one hand this can be exploited for the generation of waveguides \cite{Schmutzler2015}, but on the other hand it is detrimental when solely signal amplification is required. 
For suppression of beam deflection by background carriers the polariton condensate can be generated in a one-dimensional photonic wire \cite{Gao2012,Wertz2012,Fischer2014b}. However, this approach does not allow for a modification of the flow direction at will and the maximum gain achievable is limited due to Coulomb repulsion by background carriers. 
Here, we demonstrate that polariton traps buried in a two-dimensional microcavity can be used to compensate the repulsive interaction between the condensate and background carriers. Thereby, the propagating polariton condensate is amplified without beam deflection.

\section{Experimental details}

For the experiments presented here, two different samples were studied:
First, a planar GaAs-based microcavity with a Rabi splitting of $9.5~\mbox{meV}$ which is used as a reference sample for the demonstration of polariton beam scattering mediated by background carriers. Further details on the sample can be found in Ref.~\onlinecite{Schmutzler2015}. Second, a GaAs-based microcavity sample with a Rabi splitting of $11.5~\mbox{meV}$ containing circular-shaped regions with an elongated microcavity layer used for the demonstration of amplification of a directed polariton beam without deflection by background carriers.
Here, regions where the microcavity is elongated provide an attractive trapping potential of $4-5~\mbox{meV}$ depending on the exciton-cavity detuning.
Further details on the sample fabrication can be found in Ref.~\onlinecite{Winkler2015}.

\begin{figure}[!htp]
\centering
\includegraphics[width=0.75\linewidth]{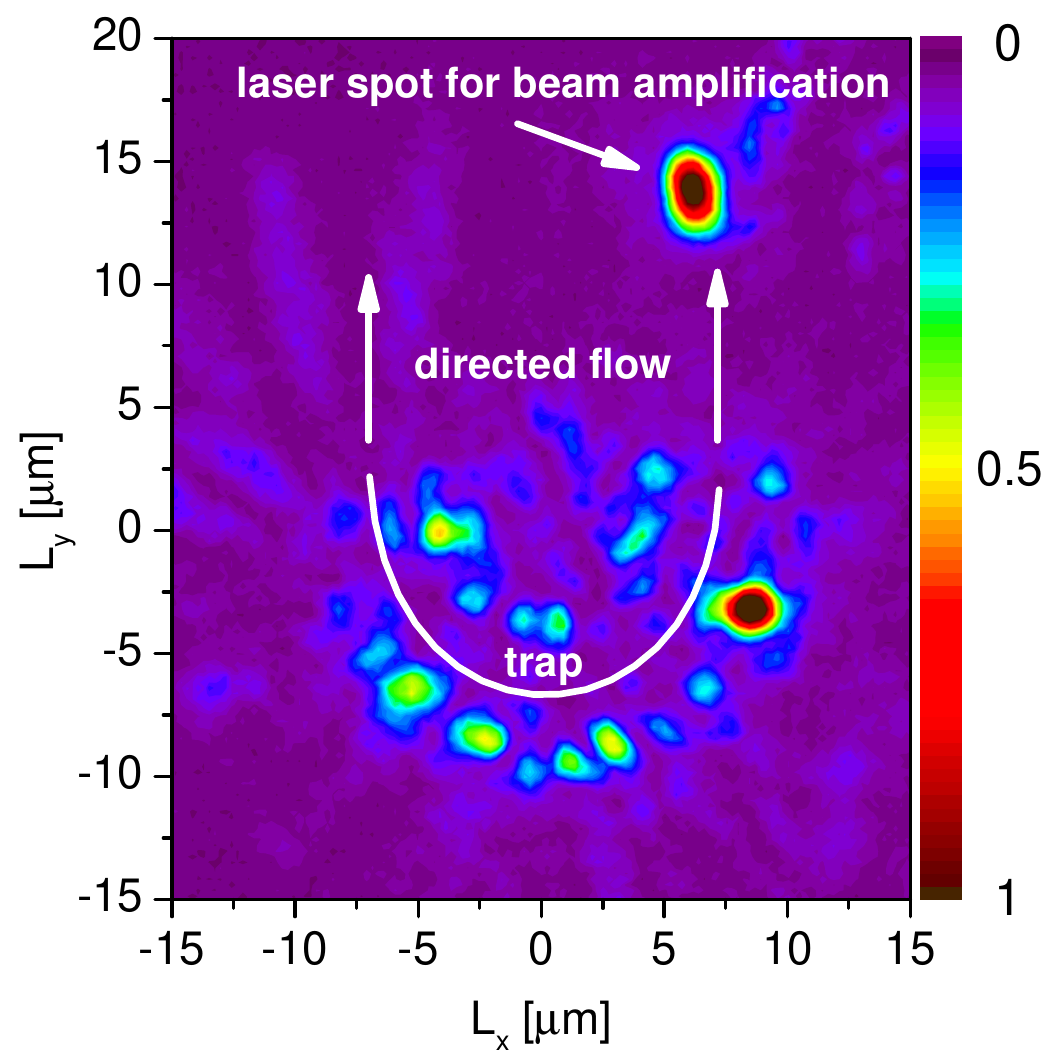}
\caption{(Color online) Excitation pattern applied for the experiments. A laser spot consisting of two concentric semicircles is imprinted onto the sample for the generation of a directed polariton beam. A repulsive potential mediated by background carriers created by a Gaussian laser spot allows for the realization of beam amplification experiments.}
\label{Fig1}
\end{figure}

The experimental setup is analogous to the one used in Ref.~\onlinecite{Schmutzler2015}:
a directed polariton beam is realized under non-resonant excitation using a spatial light modulator (SLM), which allows for excitation with arbitrarily shaped laser spots \cite{Assmann2012}. A second  Gaussian shaped laser beam, originating from the same femtosecond-pulsed Titanium-Sapphire laser (repetition rate $75.39~\mbox{MHz}$) is focused onto the sample under normal incidence with a full width at half maximum (FWHM) of approximately~$2~\upmu\mbox{m}$. This beam injects additional background carriers in the sample which form excitons that act as a gain medium for the polariton beam.

\section{Results and discussion}
We have performed the amplification experiment in the planar microcavity reference sample first to provide unambiguous evidence for the beneficial effect of polariton traps for signal amplification.
Fig.~\ref{Fig1} shows the excitation pattern used for the  experiments. Two concentric semi-circle shaped laser spots, ca.~$10~\upmu\mbox{m}$ and $15~\upmu\mbox{m}$ in diameter, respectively, are imprinted on the sample using the SLM. At the location of the excitation laser spot, background carriers are created, which provide a repulsive potential for the polariton condensate \cite{Vladimirova2010,Wertz2010,Schmutzler2014}. The purpose of this excitation pattern is to realize a trapping of the polariton condensate between the semi-circle shaped background carrier reservoirs and to generate a directed polariton beam leaving the trap through the open apertures\cite{Schmutzler2015} like in a chicane. In addition, a reservoir of background carriers is imprinted on the sample roughly $12~\upmu\mbox{m}$ away from the right aperture of the polariton chicane by the Gaussian laser spot. The exciton-cavity detuning at the investigated sample position is $\delta=-24.7~\mbox{meV}$, which corresponds to an excitonic fraction of the LP of $3-4\%$ in the range of momenta studied here.

\begin{figure*}[!t]
\centering
\includegraphics[width=\linewidth]{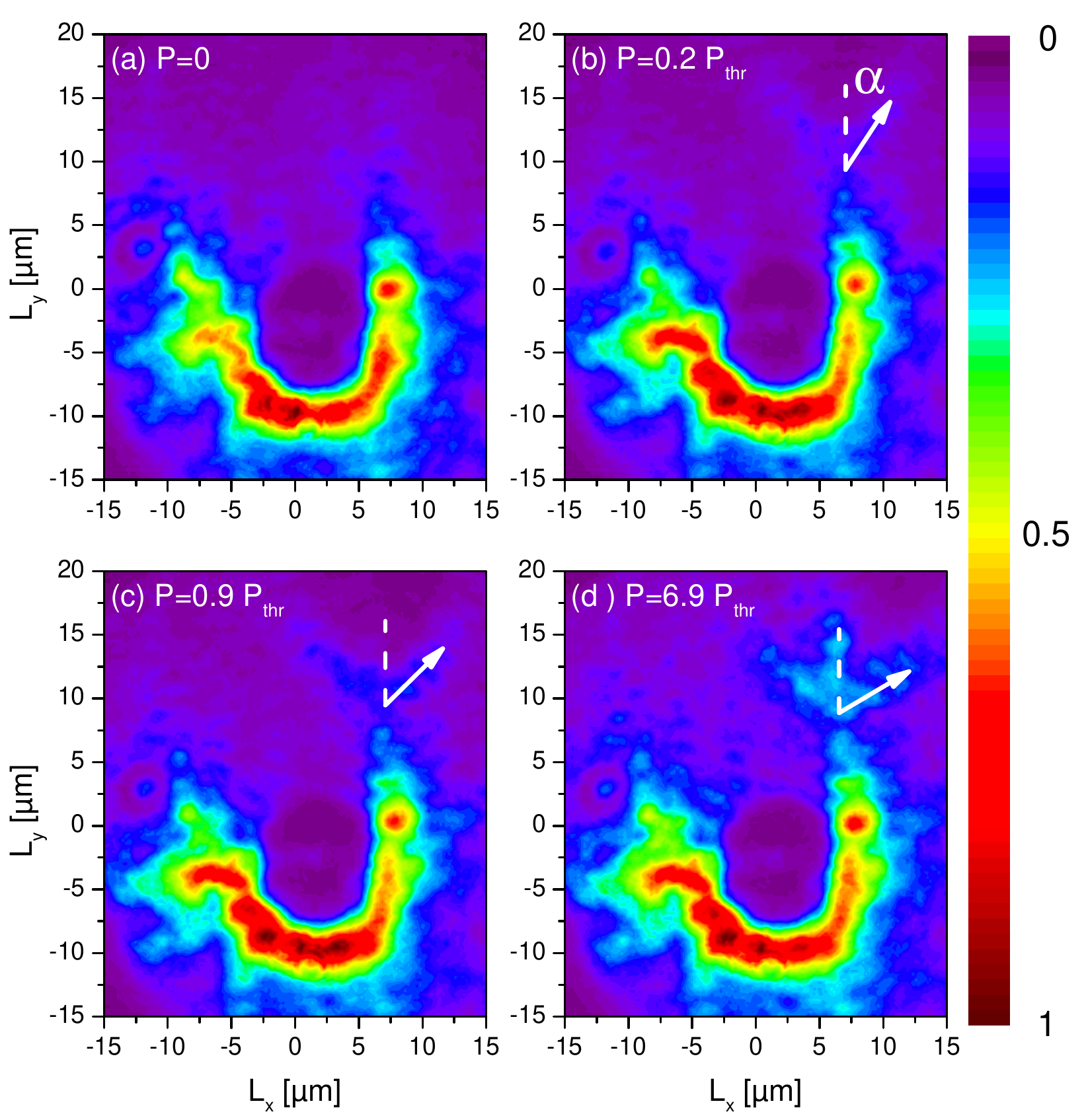}
\caption{(Color online) Polariton condensate distribution for different excitation power levels of the Gaussian laser spot. (a) Without scattering potential the polariton beam leaves the trap without further deflection in $L_\mrm{y}$-direction. \mbox{(b)-(d)~With~applied} repulsive potential pronounced scattering of the polariton beam is observed.}
\label{Fig2}
\end{figure*}

The polariton beam trajectories for the excitation pattern presented in Fig.~\ref{Fig1} are depicted in Fig.~\ref{Fig2} for four different excitation power levels of the Gaussian laser beam. Fig.~\ref{Fig2}~(a) refers to a blocked Gaussian laser beam. Here,  an undisturbed polariton beam leaving the right aperture of the trap can be observed. Fig.~\ref{Fig2}~(b)-(d) show the  polariton trajectories when the Gaussian laser beam is turned on. On the one hand, there is clear evidence for an amplification of the polariton beam by the background exciton reservoir. On the other hand, however, even at low power of the Gaussian laser spot, the polariton beam is scattered at the background carrier reservoir and two scattered polariton beams, symmetric with respect to the incoming polariton beam, can be seen. Further, the scattering angle $\alpha$ increases with excitation power of the Gaussian laser beam due to larger density of background carriers which causes an increase of the scattering potential height and lateral size. 

An evaluation of the scattering geometry reveals a scattering angle of $30^\circ$ even for low excitation powers of $P=0.2~P_{\mrm{thr}}$ [Fig~\ref{Fig2}~(b)] of the Gaussian laser spot, where $P_{\mrm{thr}}$ denotes the condensation threshold under excitation with the Gaussian laser spot only. For lower powers, no significant deflection or amplification of the polariton beam has been observed. With increasing potential height the scattering angle grows rapidly and exhibits a saturation value of roughly $60^\circ$ [Fig~\ref{Fig2}~(d)].

For the purpose of signal amplification by background excitons it would be beneficial to be able to switch off the repulsive interaction between polariton condensates and background carriers to generate gain without perturbation of the polariton beam. The repulsive potential mediated by background carriers is known from photoluminescence measurements to be on the order of $5~\mbox{meV}$ in the kind of GaAs-based microcavity used here \cite{Schmutzler2014,Schmutzler2015}. In the following we will locate the background carriers providing gain for the polariton beam in an attractive potential environment of the same order of magnitude. As described before, this can be realized by polariton-traps consisting of a circular-shaped region with elongated cavity layer. Here, the attractive potential is mediated by the photonic fraction of the LP. 

\begin{figure*}[!t]
\centering
\includegraphics[width=\linewidth]{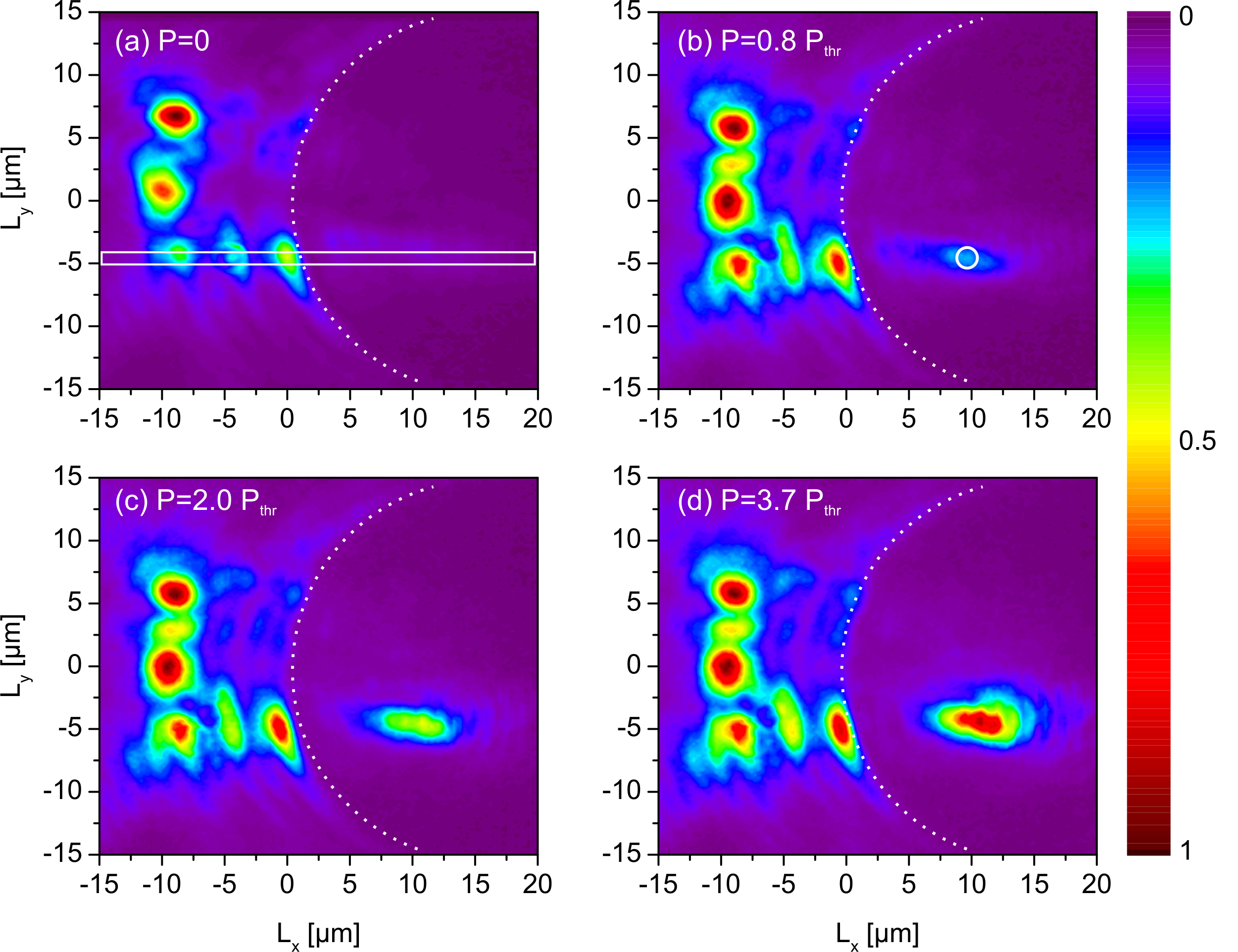}
\caption{(Color online) Generation of a directed polariton beam in the vicinity of a polariton trap. The diameter of the trap amounts to $d=30~\upmu\mbox{m}$ and is indicated by the white dashed line. The location of the Gaussian laser spot is denoted by the white circle in panel (b). (a) Polariton condensate distribution with Gaussian laser spot turned off. \mbox{(b)-(d) Polariton} condensate distribution with Gaussian laser spot at different excitation power levels. Spectral resolution is provided by an interference filter centered at $1570.2~\mbox{meV}$ with a FWHM of $2~\mbox{meV}$.}
\label{Fig3}
\end{figure*}

Fig.~\ref{Fig3} shows two-dimensional polariton distributions occurring in this experimental geometry. For comparability with the reference sample, a sample position with similar LP excitonic fraction of $4\%$ at zero momentum and the same excitation pattern as used for the planar microcavity reference sample [Fig.~\ref{Fig1}] are chosen.
The polariton beam is generated outside of the polariton trap, which is $d=30~\upmu\mbox{m}$ in diameter [indicated by the white dashed line in Fig.~\ref{Fig3}]. The beam is partly transmitted into the polariton trap [Fig.~\ref{Fig3}~(a)] and can be amplified by adding a Gaussian laser spot within the trap [indicated by the solid circle in Fig.~\ref{Fig3}~(b)]. As becomes evident from the experimental data [Fig.~\ref{Fig3}~(b)-(d)], the polariton beam is largely amplified with increasing excitation power of the Gaussian laser spot without deflection of the polariton beam.

\begin{figure*}[!t]
\centering
\includegraphics[width=\linewidth]{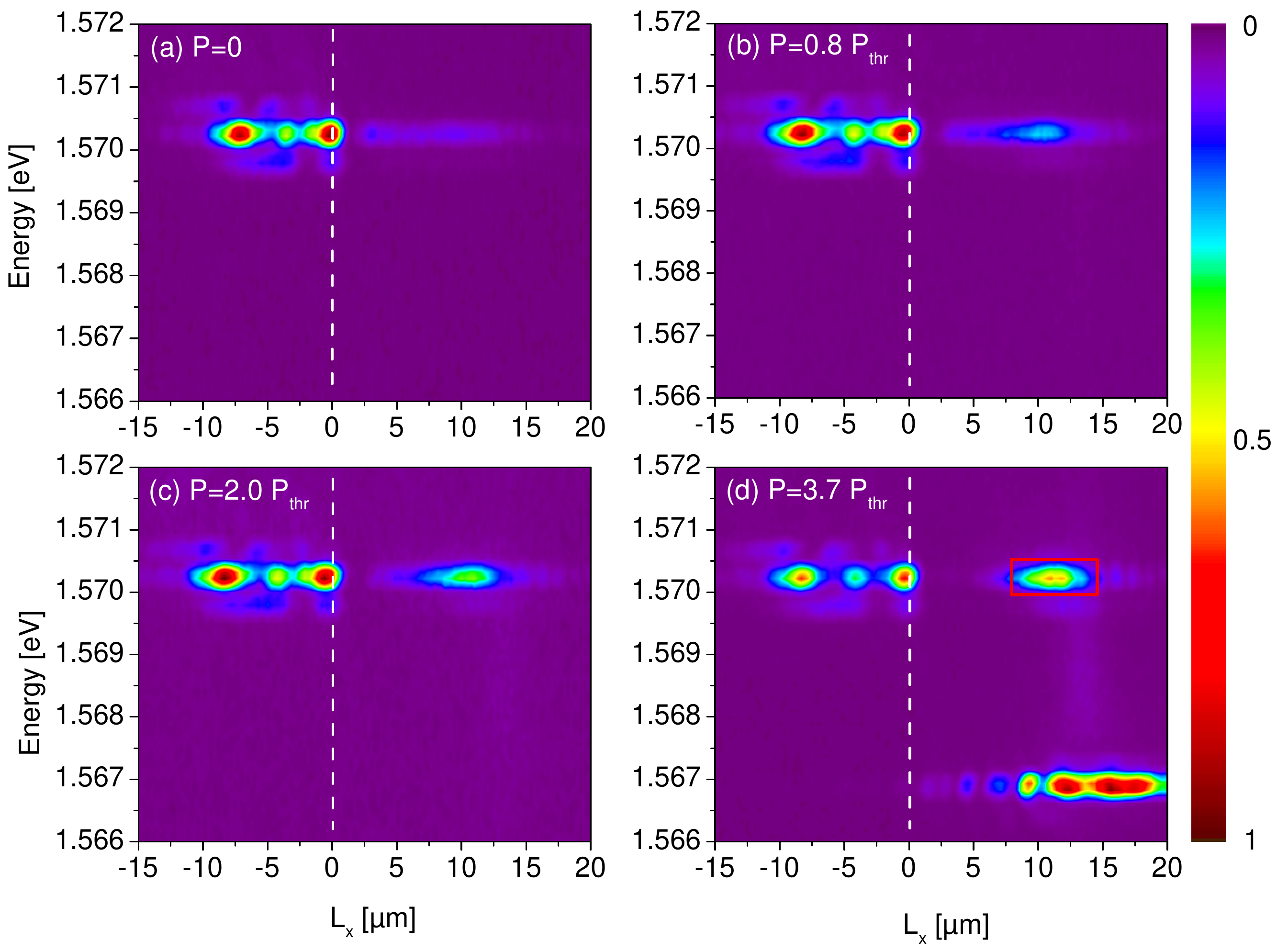}
\caption{(Color online) Spectrally resolved cross-section of a sample region $1~\upmu\mbox{m}$ in diameter centered around the polariton beam propagating in $L_x$-direction as indicated in Fig.~\ref{Fig3}~(a) by the white box. The white dashed line indicates the boundary of the polariton trap. Excitation power levels of the Gaussian laser spot are the same as in the corresponding panels of Fig.~\ref{Fig3}. The red box indicates the integration area on the CCD chip used for the analysis of the gain presented in Fig.~\ref{Fig5}~(a).}
\label{Fig4}
\end{figure*}

For a more detailed analysis we have investigated an energy resolved cross-section [indicated by the white rectangular box in Fig.~\ref{Fig3}~(a)] of the propagating polariton beam [Fig.~\ref{Fig4}]. Clearly, the polariton beam flows ballistically over the polariton trap without scattering into lower energy polariton states of the trap [Fig.~\ref{Fig4}~(a)]. This finding agrees with earlier observations where inefficient scattering between delocalized 2D-states and confined states in the polariton trap has been observed \cite{Paraiso2009}.

When the Gaussian laser beam is turned on with power levels near the condensation threshold of the trapped polariton states, the polariton flow within the polariton trap gets amplified significantly [Fig.~\ref{Fig4}~(b)]. 
Interestingly, even for an excitation power of the Gaussian laser spot above the condensation threshold of the polariton trap, solely the delocalized 2D polariton beam is amplified without  population of trapped polariton states [Fig.~\ref{Fig4}~(c)]. 
Only for the case of an excitation power level chosen significantly above condensation threshold, the trapped states are dominantly populated [Fig.~\ref{Fig4}~(d)].

\begin{figure}[!t]
\centering
\includegraphics[width=\linewidth]{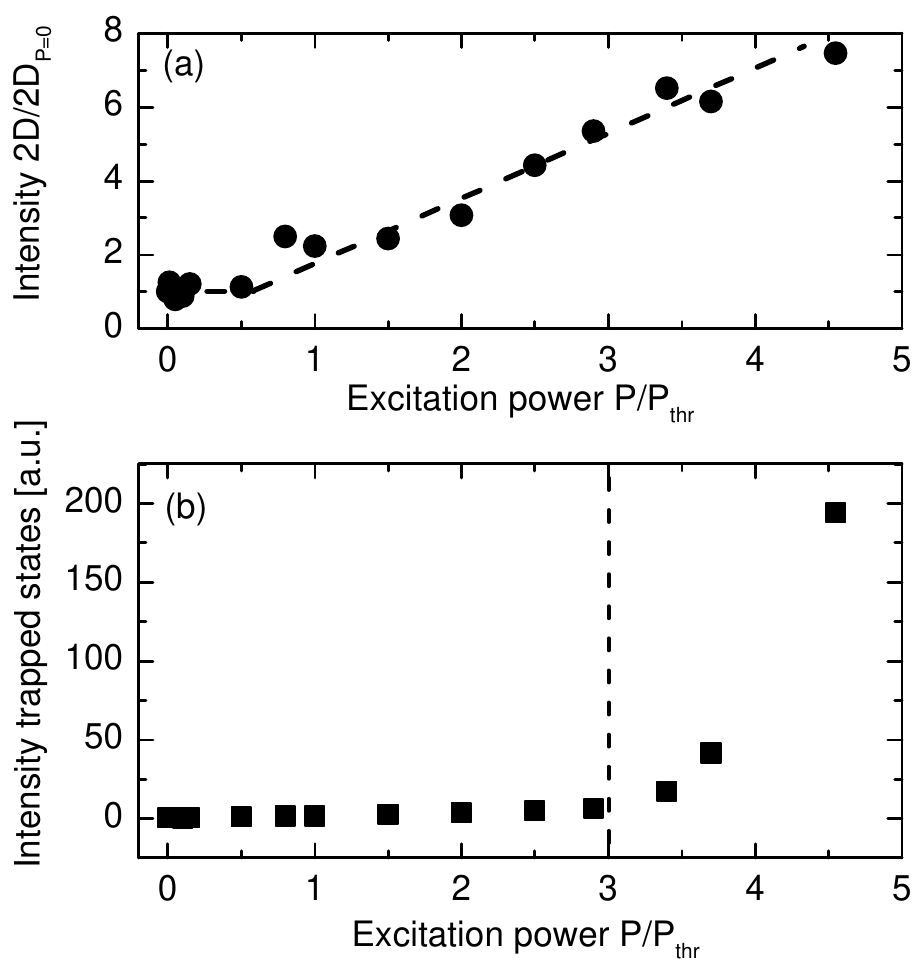}
\caption{(a) Number of counts within the area indicated by the red box in Fig.~\ref{Fig4}~(d) normalized by the number of counts without excitation of the Gaussian laser spot. 
The black dashed line is a guide to the eye. (b) Signal intensity arising from all trapped polariton states. Excitation power is normalized to the polariton condensation threshold of the trapped polariton states under excitation with the Gaussian laser spot only.}
\label{Fig5}
\end{figure}

For an evaluation of the amplification of the polariton beam by the Gaussian laser spot, we have integrated the number of counts of the CCD camera around the location of the Gaussian laser spot in the spectral region of the free propagating polariton condensate [indicated by the red box in Fig.~\ref{Fig4}~(d)]. By normalization to the number of counts without excitation of the Gaussian laser spot, the gain of the signal can be estimated. For low excitation power levels below $50\%$ of the condensation threshold of the trapped polaritons, no significant increase of the 2D polariton beam is observed [Fig.~\ref{Fig5}~(a)]. At higher excitation power levels a linear response can clearly be seen, allowing for the realization of a signal gain up to a factor of $7$. Furthermore, condensation of the trapped states sets in only at an increased excitation power of $3~P_{\mrm{thr}}$ [Fig.~\ref{Fig5}~(b)] with respect to the situation when no polariton beam is injected into
 the trap.
To reveal further insights into the interplay between background carriers, trapped polaritons and the polariton beam we have additionally performed one-dimensional numerical simulations. 

\begin{figure*}
\centering{}
\includegraphics[width=\linewidth]{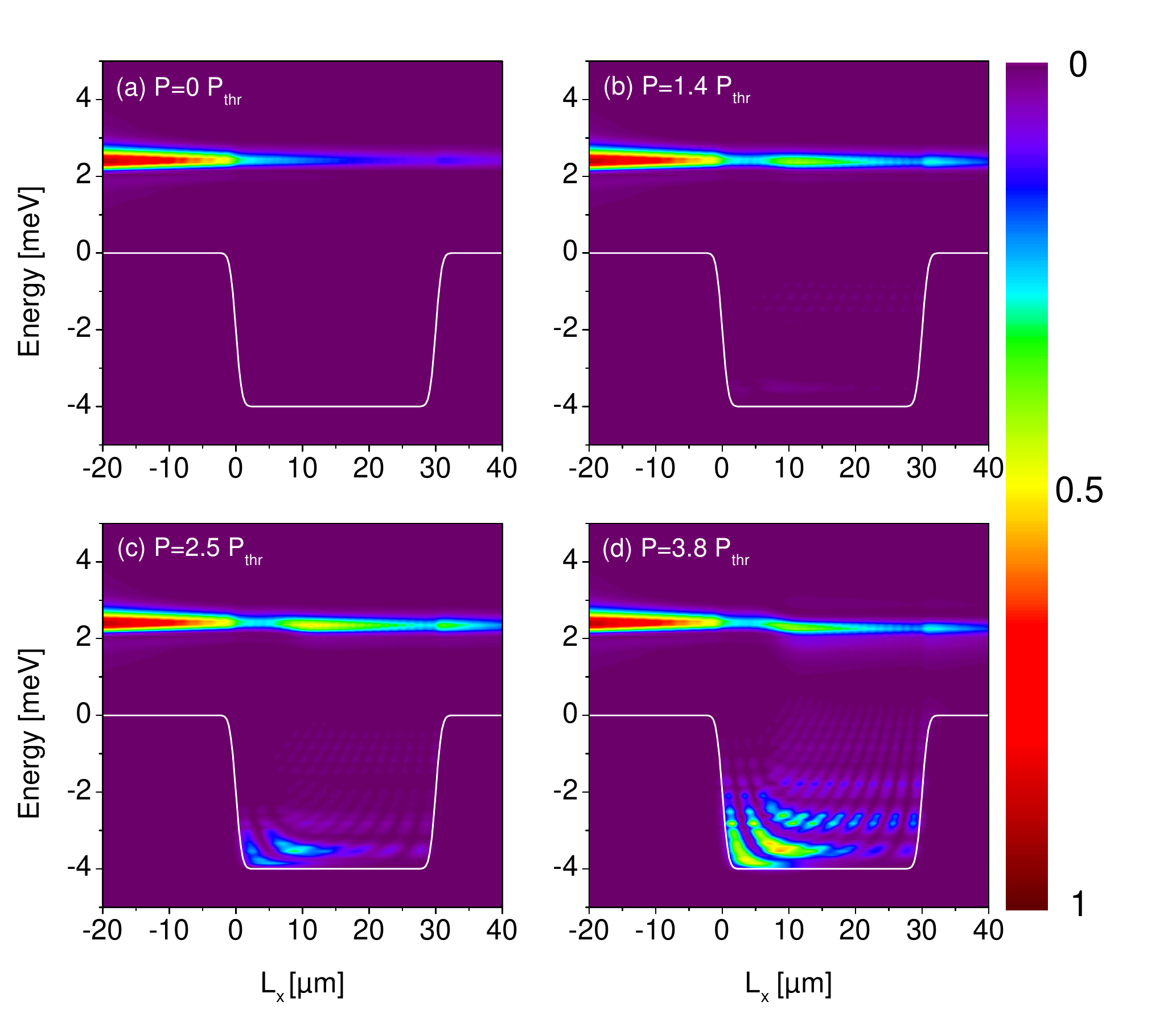}
\caption{Calculated spatial resolved polariton distribution for four different target densities. Target densities are normalized to condensation threshold of the trapped states without impinging polariton beam. The potential geometry of the trap is indicated by the white line.}
\label{theoryfig1}
\end{figure*}
%
%
Our theoretical approach is based on an extended Gross-Pitaevski equation for the coherent polariton field $\Psi(x,t)$, including the interaction with incoherent excitons. For the latter we consider a two-reservoir model, comprising ``active" excitons $n_A(x,t)$, able to undergo stimulated scattering into the coherent condensate, and ``inactive" excitons $n_I(x,t)$, which do not fulfill the energy and momentum conservation for this process \cite{Lagoudakis2011}. For legibility, the spatial and temporal dependence is omitted from here on.
The coupled dynamics of polaritons and excitons in real space and time domain are given by \cite{Schmutzler2015}:
\begin{eqnarray}
\mathrm{i} \hbar \frac{\partial}{\partial t}\Psi & = & \left(\mathbb{H} - \mathrm{i} \left(\gamma_p-\frac{\gamma}{2} n_A\right)+ V_\mathrm{trap} \right) \Psi + \alpha_1 \vert \Psi \vert^2 \Psi \nonumber \\
& + &  \left(\alpha_2 n_A + \alpha_3 n_I \right) \Psi  - \mathrm{i} \Lambda \left( n_A+n_I\right) \mathbb{H} \Psi \mbox{,}\\
\mathrm{i} \hbar \frac{\partial}{\partial t}n_A & = & - \mathrm{i}\gamma \vert \Psi \vert^2 n_A - \mathrm{i} \gamma_A n_A + \mathrm{i}\tau n_I \mbox{,}\\
\mathrm{i} \hbar \frac{\partial}{\partial t}n_I & = & - \mathrm{i} \tau n_I - \mathrm{i} \gamma_I n_I\mbox{.}
\end{eqnarray}
Here, $\mathbb{H}$ denotes the free particle Hamiltonian $\mathbb{H} = -\frac{\hbar^2}{2 m_p} \frac{\partial^2}{\partial x^2}$ with the effective mass $m_p = 0.2$ meV$ \mathrm{ps}^{2} \upmu \mathrm{m}^{-2}$. The repulsive polariton-polariton interaction is included by $\alpha_1 = 0.0024$ meV$\upmu \mathrm{m}$, the interaction between polaritons and reservoir by $\alpha_2 = 0.008$ meV$\upmu \mathrm{m}$ and $\alpha_3 = 0.008$ meV$\upmu \mathrm{m}$ for active and inactive excitons, respectively. Polaritons decay with the rate $\gamma_p = 0.1$ meV, and the loss of active and inactive excitons amounts to $\gamma_A = 0.01$ meV and $\gamma_I = 0.0013$ meV, respectively. The condensate is amplified by active excitons in terms of a stimulated scattering process with $\gamma = 0.012$ meV$\upmu$m. The active reservoir, however, is fed by inactive excitons, turning into active excitons with the constant rate $\tau = 0.3$ meV. A relaxation mechanism is included by the term $\Lambda = 2.5 \cdot 10^{-5} \upmu$m. The initial density of propagating polaritons is  $10^{6} \mathrm{cm}^{-1}$.  
Further details on the theoretical model and on the condensation process for pulsed excitation are given in Ref.~\onlinecite{Schmutzler2015}.

In the simulations, an initial polariton condensate is propagating with a momentum $k = 1.5$~$\upmu \mathrm{m}^{-1}$ corresponding to an energy of $2.4~\mbox{meV}$ above the polariton ground state against a potential trap. The size of the photonic potential $V_\mathrm{trap}$ is $30~\upmu\mbox{m}$ with a depth of $4~\mbox{meV}$ in accord with the experimental situation. 
In addition, an inactive exciton reservoir with a narrow Gaussian shape - the target - is applied inside the trap which corresponds to the Gaussian laser beam in the experiment.
\begin{figure}
\centering{}
\includegraphics[width=\linewidth]{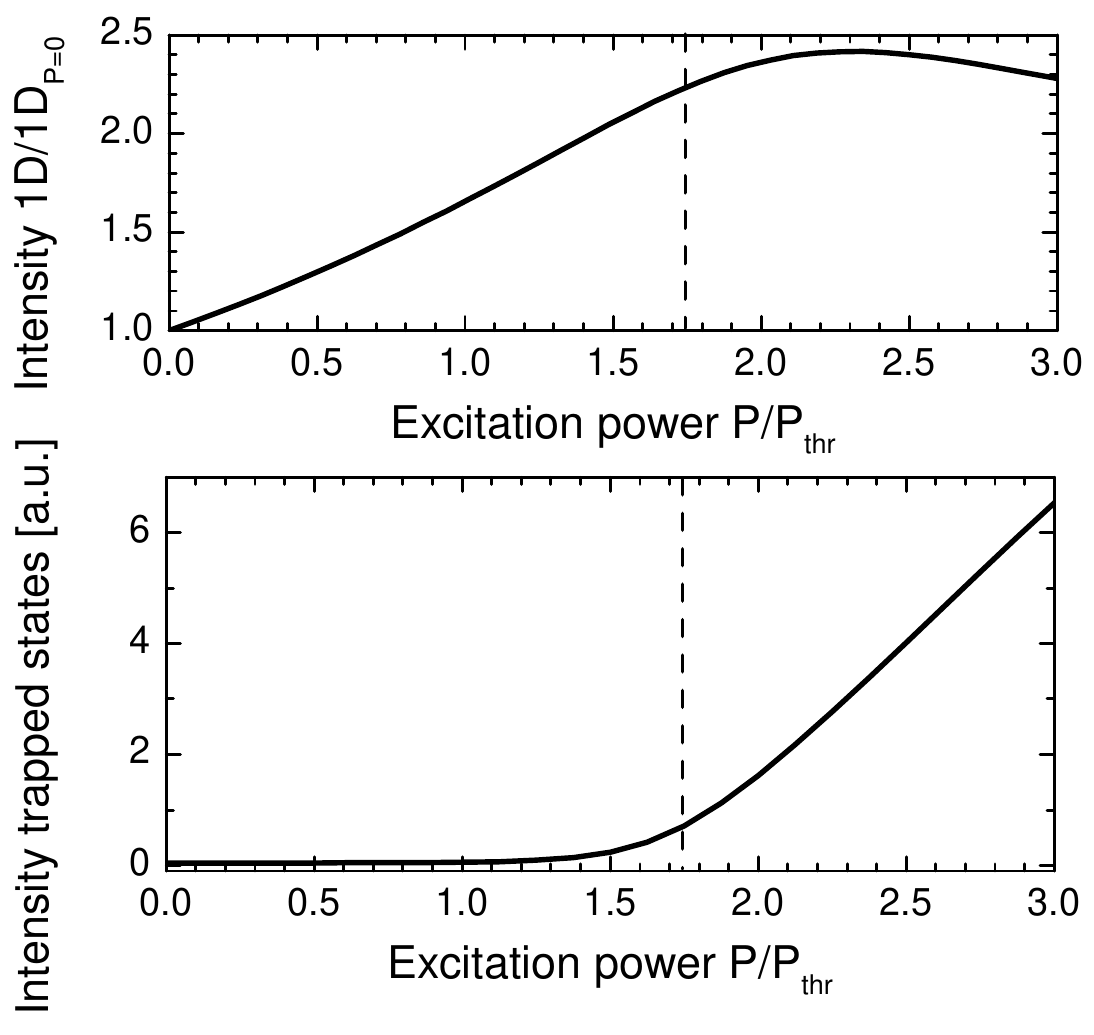}
\caption{Shown is the time integrated density of the propagating (a) and trapped polariton condensate (b) dependent on the target power. The target power is normalized to the condensation threshold without any competitive polariton propagation ($\mathrm{P_{thr}}$). The increased condensation threshold in presence of a propagating condensate is indicated by black dashed lines.}
\label{theoryfig2}
\end{figure}
The time integrated polariton density after propagation through the whole trap is shown in Fig.~\ref{theoryfig1}. The results are shown for different target densities:
without applied target, the condensate propagates ballistically through the trap without any relaxation into trapped polariton states  [Fig.~\ref{theoryfig1}~(a)] as observed in the experiment [Fig.~\ref{Fig4}~(a)].  If the target is switched on, the propagating polariton condensate is amplified due to stimulated scattering of background carriers in the active reservoir [Fig.~\ref{theoryfig1}~(b)-(d)]. 
The amplification of propagating polaritons and the time-integrated density of the trapped condensate intensity are shown in Fig.~\ref{theoryfig2}~(a) and (b), respectively. 
In agreement with the experimental observation, the condensation threshold for the trapped states is shifted to higher power levels when a propagating polariton condensate is injected into the trap [Fig.~\ref{theoryfig2}~(b)].
Furthermore, our calculations also indicate a regime of monotonously growing amplification of the propagating polariton condensate up to \mbox{$2.3$ $\mathrm{P_{thr}}$} [Fig.~\ref{theoryfig2}~(a)]. 
However, above $2.3$ $\mathrm{P_{thr}}$ a regime of decreasing amplification can be seen, which has not been observed in the experiment.
Remarkably, although our model is simple, the main features of our experiment, namely the shift of the condensation threshold of the trapped states to higher power levels as well as the amplification of the propagating condensate are recovered which allows for the following interpretation of our observations:
the relaxation dynamics of background carriers are modified due to the presence of a polariton beam penetrating into the trap.
Here, scattering into untrapped states depopulates the background carrier reservoir, which  causes an increase of the condensation threshold of the trapped polaritons compared to the situation when no macroscopic population of untrapped states is present. In the simulations, the observed decrease of amplification of the polariton beam indicates a competition between spontaneous condensation process into trapped states and the amplification of the polariton beam occurring for high excitation power levels. However, the available power in the experiment did not allow for approaching this regime which should be investigated in future work.

%
For operation as an amplifier a trade-off between maximum signal gain and a minimum generation of potentially unwanted polariton population in the trap has to be chosen, which lies in a range between $0.8~P_{\mrm{thr}}$ and $3~P_{\mrm{thr}}$ in the experiments presented here.
Additionally, our approach might also be considered as a simple switching device as the propagating condensate is able to turn off the population of trapped states in the power range of $1-3~P_{\mrm{thr}}$.

In conclusion, a novel type of polariton amplifier has been presented. Using polariton traps, the repulsive interaction between background carriers and polariton condensates can be turned off allowing for polariton beam amplification without distortion. This finding might be of relevance for long range signal transmission in polaritonic logic circuits.

\section{Acknowledgements}
The Paderborn and Dortmund group acknowlegde support by the collaborative research center TRR 142 of the DFG. The Paderborn group further thanks for computing time at PC2 and support through the DFG Heisenberg programme. The group at W\"urzburg University acknowledges support from the State of Bavaria and the assistance of A. Schade, J. Gessler and M. Emmerling during sample preparation. M.B. acknowledges support from the Russian Ministry of Science and Education (contract number 14.Z50.31.0021). 

%

\end{document}